\documentclass[prb,aps,graphicx,color,float]{revtex4}
\usepackage{epsfig}
\usepackage{bm}

\begin{document}

\preprint{}

\newcommand{\stef}[2]{$\blacktriangleright${\sc round #1:}{\em #2}$\blacktriangleleft$}

\title{Quantum Phase Transition between (Luttinger) Liquid and Gas of Cold Molecules}

\author{K. T. Law and D. E. Feldman}
\affiliation{Department of Physics, Brown University, Providence,
Rhode Island 02912, USA}

\begin{abstract}

We consider cold polar molecules confined in a helical optical lattice similar to those used in holographic microfabrication. An external electric field polarizes molecules along the axis of the helix. The large-distance inter-molecular dipolar interaction is attractive but the short-scale interaction is repulsive due to geometric constraints and thus prevents collapse. The interaction strength depends on the electric field. We show that a zero-temperature second-order liquid-gas transition occurs at a critical field. It can be observed under experimentally accessible conditions.
\end{abstract}

\pacs{71.10.Pm, 64.70.F-, 03.75.Lm}

\maketitle

At zero temperature most substances exist in the form of solids. The only exception is Helium which undergoes quantum melting at a critical pressure. On the other hand, zero-temperature liquid-gas transitions have never been observed.
Indeed, at absolute zero any system must be in its ground state and condensed phases have lower energy than gases.
Driving a quantum liquid into a zero-temperature gas state would be possible, if one could control inter-atomic forces. 
At sufficiently weak inter-atomic interactions, a many-particle bound state (i.e., liquid or solid) would cease to exist and a gas would form instead. While this cannot be accomplished with conventual materials, recent progress in the field of cold dilute gases opens a possibility to tailor a wide range of Hamiltonians with tunable parameters. In this Letter we show that a quantum liquid-gas transition can be observed in a cold gas of polar molecules confined in an optical lattice.

Experiments with cold gases have already allowed the observation of Bose-Einstein condensation, BCS superfluidity and Mott localization \cite{review}. It was proposed that cold gases can serve as realizations of other analogies of electronic matter such as superconductors with p-wave pairing \cite{pwave} and quantum Hall states 
\cite{QHE}. Besides, several new states of matter with different broken symmetries and/or soft modes were predicted in cold atom systems. We address a rather different situation.
Liquids and gases have the same symmetry and their only difference is the density: A gas fills all available volume while the density of a liquid is determined by inter-molecular interactions. 

While cold gases do not represent true ground states of alkali metals, they are highly stable due to their low density which guarantees a low probability of multi-particle recombination processes \cite{book}. A well-established way to control interaction in cold gases utilizes Feshbach resonances and 
allows changing both the strength and sign of the short-range inter-atomic potential which can be modeled as a delta-function in space. If it is repulsive the system minimizes its energy by occupying all available volume, i.e., is a gas. Attractive interactions have different effects on fermions and bosons. A Fermi gas undergoes Cooper pair formation while a Bose gas collapses into a regime with strong many-body recombination. In our case a different inter-atomic potential in needed: A liquid state can be formed, if the interaction is attractive at large scales but the short-range force must be repulsive to prevent collapse.

We demonstrate that such potential can be built from the dipole interaction of polar molecules \cite{dipole,three-body}.
The sign of the dipole interaction is certainly independent of the distance and depends only on the direction of the dipole moments. It was shown recently \cite{three-body} that the interaction sign can be made distance-dependent by driving molecules with microwave fields. However, such an approach generates many-body forces \cite{three-body}. It would be interesting to investigate if a liquid phase is possible in such a system but in this paper we focus on a simpler situation with only two-body interactions. It can be achieved by confining polar molecules in a helical optical lattice, i.e., a potential well of a helical shape. Hexagonal arrays of such helices \cite{hexagon} are among numerous periodic and aperiodic structures used in holographic microfabrication experiments. We will see below that such structures open new possibilities for tailoring
cold-atom Hamiltonians which cannot be obtained with usual optical lattices in the form of sinusoidal waves.
 
We focus on the simplest problem of this type with the lattice in the form of a single helix (Fig. 1a). It can be produced with an approach similar to Refs. \onlinecite{hexagon} as discussed in the Appendix. An external electric field polarizes molecules along the axis of the helix. In zero field, molecules have zero angular momentum and hence zero average dipole moment while in strong fields the average dipole moment $p$ can reach values of the order of Debyes. At large inter-molecular distances
the dipole interaction $V=p^2(1-3\cos^2\theta)/r^3$ is attractive ($\theta=0$). At short distances it becomes repulsive, if the angle $\gamma$ between the tangent to the helix and its axis exceeds the magic angle $\cos^{-1} (1/\sqrt{3})$. The distance dependence of the interaction may include multiple maxima and minima for $\gamma\approx\pi/2$. In this paper we focus on smaller $\gamma$ so that the interaction dependence on the distance $s$ along the helix has a simpler shape shown in Fig. 1b. The potential resembles the Lennard-Jones interaction used in models of thermal liquid-gas transitions and we show that a transition between a Luttinger liquid and Fermi or Tonks-Girardeau gas occurs at a critical value of the dipole moment $p$.

The paper is organized as follows: we first formulate the model. Next, we find its phase diagram with a variational method. We support the variational calculation by a proof that for a class of inter-molecular potentials there is a phase transition between a monoatomic gas and a liquid. Then we address the conditions for the experimental observation of the transition.

We consider $N\gg 1$ particles of mass $m$ confined in a helical potential well of radius $R$ and pitch $d\sim R$. We assume that the wave function is confined in a region of the width $\sim w\ll R$ around the helix (Fig.1a). The adiabatic approximation applies and at low temperatures $T\ll \hbar^2/(mw^2)$ the problem reduces to a one-dimensional model with the Hamiltonian 

\begin{equation}
\label{1}
H=-\sum_i\frac{\hbar^2}{2m}\frac{\partial^2}{\partial s_i^2}+\sum_{i>j}V(s_i-s_j),
\end{equation}
where $s_i$ is the length of the helix between particle number $i$ and a reference point on the helix,
$V(s)$ the dipole interaction. 
Repulsive short-range interaction keeps particles apart. Hence, in 1D the particle statistics is unimportant. In what follows we focus on non-identical particles with $s_{n+1}>s_n$ for all $n$. The ground state energy is independent of statistics and the wave functions of identical bosons and fermions can be obtained from the case of non-identical particles by symmetrization or antisymmetrization.


\begin{figure}
\epsfig{file=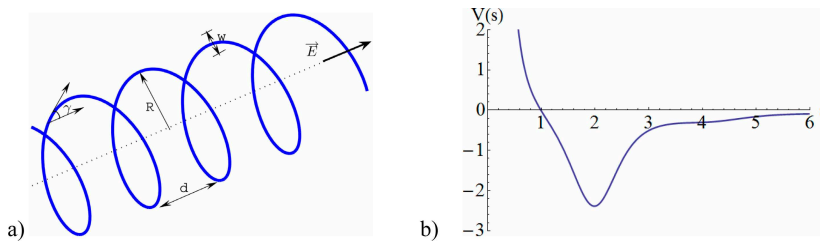, scale=0.6} 

\caption{a)~Helical lattice. b)~Potential  $V(s)$ in units of  ${p^2}/{({\pi}R)^3}$ as a function of distance $s$ in units of $\pi R$. $\gamma=\arctan 2$.}
\label{fig1}
\end{figure} 

As the first step, we use the self-consistent harmonic approximation (SCHA), well-known in the theory of disordered classical systems \cite{var1,var}. A better approximation will be considered in the end of the paper. Note that a method, similar to SCHA, has been used in a related problem \cite{ek}. We interpret the Euclidean action $S$ as a classical Hamiltonian and $\hbar$ as the temperature.
We consider a trial distribution function of the form $P\sim\exp(-S_0/\hbar)$, where the trial action 

\begin{equation}
\label{2}
S_0=\frac{1}{2}\int dt [\sum_{k=1}^
{N}{m\dot s_k^2}+\sum_{k=1}^{N-1}{K(s_{k+1}(t)-s_k(t)-h)^2}]=
\int \frac{d\omega}{2\pi}\sum_{k,l=1}^N G_{kl}(\omega)s_k(\omega)s_l^*(\omega)+
\int dt Kh(s_1-s_N+[N-1]h/2)
\end{equation}
contains the variational parameters $h$ and $K$; $h$ is related to the average length of the $N$-particle chain and
$K$ describes the fluctuations of the particle positions $s_k$. We need to minimize the variational free energy
$E(K,h)=[\langle S-S_0\rangle_0 -\hbar\ln Z_0]/\tau,$
where $\langle\dots\rangle_0$ denotes the average with respect to the trial distribution function $P$, the trial partition function $Z_0=\int \Pi ds_k \exp(-S_0/\hbar)$, and the time $\tau\rightarrow\infty$ is the size of the integration domain. The self-consistent harmonic approximation provides analytic results for potentials $V(s)$ in the form 
of the sums of exponentials. Thus, we approximate the potential represented in Fig. 1b by the Morse potential \cite{LL},
$V(s)=A[\exp(-2\alpha [s-h_0])-2\exp(-\alpha [s-h_0])]$, which has a similar shape. Since the interaction rapidly decreases at large distances, we keep only the interaction between neighboring particles ($k$ and $k+1$).

The variational calculation includes three steps: 1) we change the variables $s_k\rightarrow s_k+kh$ so that $h$ drops out of Eq. (\ref{2}); 2) calculate $E(K,h)$; and 3) find the minimum of $E(K,h)$. Step 2) requires the calculation of 
the correlation function \cite{footnote} $\langle s_n(\omega)s^*_m(\omega)\rangle=\hbar\hat G^{-1}_{nm}(\omega)/2$. The matrix $G_{nm}(\omega)$ differers from a tridiagonal Toeplitz matrix \cite{Toeplitz} only by the values of two matrix elements.  Its inverse can be found analytically using the same method as for tridiagonal Toeplitz matrices \cite{Toeplitz}. In the large $N$ limit, step 2) yields
$E(K,h)/[N\tau]=({\hbar}/{\pi})\sqrt{{K}/{m}}+A\left\{x^2\exp[{4\alpha^2\hbar}/{\pi\sqrt{Km}}]
-2x\exp[{\alpha^2\hbar}/{\pi\sqrt{Km}}]\right\},$
where $x=\exp(-\alpha [h-h_0])$. After the minimization with respect to $x$, the energy per particle $E(K,h)/[N\tau]$ reduces to 
$\epsilon(K)=(\hbar/\pi)\sqrt{(K/m)}-A\exp[-2\alpha^2\hbar/\pi\sqrt{Km}]$. The energy $\epsilon(K)$ has a local minimum at $K=0$, where $\epsilon(0)=0$. At large $A$ another minimum at $K=K_{\rm min}$ is possible. The phase transition into the state corresponding to that minimum occurs when $\epsilon(K_{\rm min})=0$. Solving the latter equation together with $d\epsilon(K_{\rm min})/dK=0$ we find a phase transition at $A_c=2e\alpha^2\hbar^2/[m\pi^2]$.

Thus, at small $A$ we get $K=0$ and $h=\infty$, i.e. the system is a gas. $K$ and $h$ are finite at large $A$. This means a finite volume at zero external pressure, i.e. a condensed state. In 1D this cannot be a solid and the calculation of the correlation function $\langle(s_{n+k}(t)-s_{n}(t)-kh)^2\rangle\approx \frac{\hbar\ln k}{\pi\sqrt{Km}}$ at $k\gg 1$ shows that the particles form a Luttinger liquid \cite{Luttinger}. 

Our problem is connected with the physics of atoms confined in Carbon nanotubes. If one ignores the periodic potential created by Carbon atoms then a model, similar to ours, emerges (certainly, one cannot tune the interaction between the atoms and obtain a phase transition in a nanotube). A variational study \cite{ek} predicted that an increasing interaction drives a monoatomic gas into a diatomic gas phase before a liquid state can be reached. This contradicts our findings. Below we sketch a proof that a liquid state has lower energy than a di- or multi-atomic gas and hence a multi-atomic gas cannot be the ground state. Qualitatively this reflects the fact that in a liquid every particle has bonds with two neighbors and only one bond in a diatomic molecule. Hence, one expects lower energy per particle in a liquid. The prediction of a diatomic gas is thus an artifact of the variational method (and Ref. \onlinecite{ek} admits such a possibility). 

Our proof is based on a variational upper bound for the ground state energy. We will also use that bound to improve SCHA.
We will focus on models in which only neighboring particles ($k$ and $k+1$) interact. A diatomic gas was predicted in Ref. \onlinecite{ek} for such a model and such a model is relevant for us since dipole forces rapidly decrease at large scales. We will discuss elsewhere a derivation of the statements, proven below, for systems in which all pairs of particles interact via attractive potentials with hard-core repulsion. We first demonstrate that the energy of a triatomic gas is always lower than the energy of a diatomic gas, provided that the interaction potential and the ground state wave functions are well-behaved. Then we use a similar argument to show that the energy of the liquid phase is lower than the energies of all possible multi-atomic gases. In all cases we assume that the potential energy is zero at an infinite interparticle separation $s_{k+1}-s_k=+\infty$ and impose the hard core condition $V(0)=\infty$.

Consider a system of $N$ particles in the interval $-\infty<s_k<\infty$.
The center of mass is at rest in the ground state. Hence, the wave function can be represented as $\psi_N(\Delta_1,\dots,\Delta_{N-1})$, where $\Delta_k=s_{k+1}-s_k$. The ground state configuration can be viewed as a set of molecules.
A molecule is a bound cluster of $n$ particles $k,(k+1),\dots,(k+n-1)$ 
with finite\cite{footnote0} interparticle distances $\Delta_k,\dots,\Delta_{k+n-2}$. The molecules are separated by 
infinite\cite{footnote01} intervals $\Delta_p$.  Different molecules do not interact. Hence, the ground state energy is the sum of the energies of separate molecules: One can easily see that if the ground state includes $k$ molecules with $N_1,\dots,N_k$ particles then the energy equals the sum of the ground state energies of the Hamiltonians (\ref{1}) with 
$N=N_1,\dots,N_k$.

We now show that the energy of a diatomic gas can always be decreased, if particles rearrange into trimers.
A diatomic gas may form, if two particles have a bound state with the energy $\epsilon_2<0$. The bound state wave function $\psi_2(\Delta_{1})$ is a normalized eigenfunction of the Hamiltonian $H_{12}$, 
given by Eq. (\ref{1}) with $N=2$. 
 Since the Hamiltonian $H_{12}=-\frac{\hbar^2}{m}\frac{d^2}{d\Delta_{1}^2}+V(\Delta_{1})$ is real, we can assume that $\psi_2$ is real \cite{footnote1}. 
In the diatomic gas, the energy per particle is $\epsilon_2/2$. We now demonstrate that there is a three-particle state with the energy per particle $\epsilon'<\epsilon_2/2$. The three-particle Hamiltonian $H_{123}=-\sum_{i=1}^3\frac{\hbar^2}{2m}\frac{\partial^2}{\partial s_i^2}+V(\Delta_{1})+V(\Delta_{2})$. Let us find the average energy $E$ of the trial wave function $\psi_3(s_1,s_2,s_3)=\psi_2(\Delta_{1})\psi_2(\Delta_{2})$. The structure of the trial wave functions prompts the change of variables $s_1,s_2,s_3\rightarrow \Delta_{1},\Delta_{2},s_3$. The Jacobian of this transformation  is one. Hence, 

\begin{equation}
\label{5}
E=\int d\Delta_{1} d\Delta_{2}
\psi_2(\Delta_{1})\psi_2(\Delta_{2})\left[V(\Delta_{1})+V(\Delta_{2})-\frac{\hbar^2}{m}(\frac{\partial^2}{\partial\Delta_{1}^2}+\frac{\partial^2}{\partial\Delta_{2}^2})+\frac{\hbar^2}{m}\frac{\partial^2}{\partial\Delta_{1}\partial\Delta_{2}}\right]\psi_2(\Delta_{1})\psi_2(\Delta_{2}).
\end{equation}
The integral of the last term in the square brackets reduces to the square of the integral of a full derivative,
$(\int d\Delta [d\psi_2^2/d\Delta]/2)^2=0$. The first four terms in the brackets can be represented as the sum of two two-particle Hamiltonians $H_{12}+H_{23}$. Thus, $E=2\epsilon_2$ 
and the energy per particle $E/3<\epsilon_2/2$. This shows that the energy is lower in the triatomic gas than in the diatomic gas which thus cannot be the ground state. 

A similar argument proves a general statement: Consider a system of $N$ particles in the interval $-\infty<s_k<\infty$.
Only nearest neighbors interact. Assume that a bound state exists for $n<N$ particles. Then in the ground state the system consists of no more than $n$
molecules and exactly one of those molecules contains more than one particle. Indeed, the ground state energy of the infinite system equals the sum of the energies of its molecules with zero center-of-mass velocities. Imagine that the ground state includes two multi-atomic molecules with $m$ and $(k+1)$ atoms.
Let $\psi_{m}(\Delta_{1},\dots,\Delta_{m-1})$ and 
$\psi_{k+1}(\Delta_{1},\dots,\Delta_{k})$ be the ground states of the Hamiltonian (\ref{1}) with $N=m$ and $N=k+1$ and the energies of these states be $\epsilon_m$ and $\epsilon_{k+1}$. The sum of these two energies contributes to the ground state energy $\epsilon_g$. We now use the argument of the previous paragraph to show that the energy decreases, if we substitute those two molecules with a monoatomic molecule (whose energy is 0) and a $(m+k)-$atomic molecule. Indeed,
the same calculation as in Eq. (\ref{5}) with the variational wave function 
$\psi_{m+k}=\psi_{m}(\Delta_1,\dots,\Delta_{m-1})\psi_{k+1}(\Delta_m,\dots,\Delta_{m+k-1})$ shows that 
the ground state energy $\epsilon_{m+k}$ of the Hamiltonian $H_{m+k}$ (\ref{1}) with $N=m+k$ cannot exceed
$\epsilon_m+\epsilon_{k+1}$. The wave function $\psi_{m+k}$ is not an eigenfunction of $H_{m+k}$
and hence $\epsilon_{m+k}<\epsilon_m+\epsilon_{k+1}$. Thus, there is a state whose energy 
$\epsilon_g -(\epsilon_m+\epsilon_{k+1})+(0+\epsilon_{k+m})$ is below the ground state energy $\epsilon_g$. The contradiction means that no more than one multi-atomic molecule exists in the ground state. Hence, if there were more than $n$ molecules in the ground state then at least $n$ of them would be monoatomic. In such situation the energy decreases, if we form an additional $n$-atomic
molecule from $n$ free particles. This proves that there are no more than $n$ molecules and no more than one of them is multi-atomic. If $N\gg n$ then exactly one molecule contains at least $N-n+1$ particles which can be described as a liquid.

The above discussion shows that only two possibilities exist for the ground state in the large-$N$ limit:
a monoatomic gas, if no bound states exist at all, and a liquid. Both possibilities take place at different interaction strengths $A$. At $A=0$ the system must be a gas \cite{lb}. At large $A$, SCHA yields a negative upper bound for the energy.
Hence, the monoatomic gas with its zero energy cannot be the ground state at large $A$ and a liquid-gas transition must occur at an intermediate $A$. 
SCHA is exact at small and large $A$.
Indeed, it correctly predicts zero energy in the gas phase at small $A$. At large $A$ the fluctuations of $\Delta_{k}$
are small. Hence, it is legitimate to expand the potential energy $V(\Delta)$ up to the second order which means that 
the action is quadratic and
SCHA is quantitatively valid. However, SCHA is insufficient near the phase transition. This is clear from the comparison of the
variational estimate for the transition point $A_c$ and the exact threshold $A_d$ for the formation of diatomic molecules in the Morse potential \cite{LL}. Contrary to the above proof, the SCHA result for $A_c$ exceeds $A_d=\hbar^2\alpha^2/(4m)$. Thus, a different method is needed near the transition. We try the variational ansatz of the form $\psi_N=\Pi_{k=1}^{N-1}\psi(\Delta_{k})$,
where the whole function $\psi$ is a variational parameter. From the calculation, completely analogous to Eq. (\ref{5}), we find the
variational energy $E=(N-1)E_2$, where $E_2$ is the average energy of a two-particle system in the state $\psi(s_1-s_2)$
 in the Morse potential. The lowest $E_2$ corresponds to $\psi(\Delta)$ which is the ground state in the Morse potential \cite{LL}. Hence,
$E_2=-A[1-\alpha\hbar/\sqrt{4mA}]^2$ at $A>A_d$ and $E=0$ at $A<A_d$. This improves an estimate for the transition point, $A_{c,{\rm new}}=A_d$. 
Obviously, $E$ is lower than the SCHA variational energy $\epsilon=0$ in the interval $A_c>A>A_{c,{\rm new}}$.
 The improved variational method leads to an unexpected 
prediction concerning the order of the liquid-gas transition. The size of a diatomic molecule diverges \cite{LL} as 
$s\sim 1/\sqrt{E}$ at $A\rightarrow A_d$. Hence, the size $l$ of the $N$-atomic bound state $\psi_N$ diverges according to the same law.
This means that the density of the liquid $\rho=N/l\sim\sqrt{E}\rightarrow 0$ at $A\rightarrow A_c$. In other words, the variational method predicts a second order liquid-gas transition. Second order transitions between Luttinger liquid states
with different densities are known \cite{pt} but in contrast to Refs. \onlinecite{pt} the symmetry does not change at our transition. According to the Landau theory such transitions must be first-order. 
On the other hand, examples\cite{Potts} are known of second order transitions in low-dimensional systems for which the Landau theory predicts the first order.
It would be interesting to find a rigorous description of the 1D liquid-gas critical point.

We see that the phase transition occurs when the characteristic kinetic and potential energies are of the same order of magnitude, $(\hbar\alpha)^2/2m\sim V(1/\alpha)$. For polar molecules in a helical lattice, the characteristic spatial scale
$1/\alpha\sim \pi R$. Hence, near the transition the dipole energy $p^2/(\pi R)^3$ must be of the order of the recoil energy $\hbar^2/m(\pi R)^2$. For a realistic optical lattice with $\pi R\sim 1\mu$m and the molecular mass of the order of 100 we find $p\sim 1$D at the transition. Such dipole moments are within reach. 
A difference between a liquid and gas can be detected in 
a variant of the Einstein's-boxes experiment. 
A laser beam, orthogonal to the helix, creates a potential barrier in the center of the helical lattice. A gas occupies all lattice and an approximately equal number of particles will remain on both sides of the barrier.
The volume of a liquid is much smaller than the system size far from the transition. Hence, all atoms will be found on one side of the barrier.

In conclusion, we have shown that a gas of polar molecules in a helical optical lattice can be driven by an electric field
into a Luttinger liquid state via a continuous phase transition. The gas fills all available volume while the volume of the liquid is determined by the interaction strength. 
We thank G.~P. Crawford, J. Dalibard, E. Demler, A. Kitaev and C. Salomon for useful discussions. We acknowledge the support by NSF under Grant No. DMR-0544116 and the hospitality of LPTENS (D.~E.~F.).

\begin{figure}
\epsfig{file=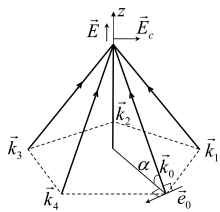, scale=0.3} 

\caption{Laser beam configuration.}
\label{fig3}
\end{figure} 

{\it Appendix.} A helical lattice can be obtained as shown in Fig. 2. A circularly polarized wave with ${\bf E}_c=E_{\rm c}(1,i,0)$
and the wave vector $k(0,0,\pm 1)$ interferes with $n$ laser beams with wave vectors ${\bf k}_m=k(\cos\alpha\cos[2\pi m/n],\cos\alpha\sin[2\pi m/n],\sin\alpha)$ and electric fields ${\bf e}_m={E_{\rm l}}(-\sin[2\pi m/n],\cos[2\pi m/n],0)$. At $n=6$ this configuration produces a periodic array of identical helices \cite{hexagon}. We focus on the mathematically simplest case of large $n$. The optical potential perceived by an atom in the $L_z=0$ state is proportional to the intensity of light\cite{book} 
$|{\bf E}|^2=2E_{\rm c}^2+[nE_{\rm l} J_1(k\rho\cos\alpha)]^2+2n E_{\rm c} E_{\rm l} J_1(k\rho\cos\alpha)\cos[kz(\pm 1-\sin\alpha)+\phi]$,
where $\phi,\rho,z$ are polar coordinates. Depending on the sign of detuning the atoms will be trapped near the intensity minimum or maximum. Both correspond to helices $kz(\pm 1-\sin\alpha)+\phi=\pi n;\rho={\rm const}$.


\begin{thebibliography} {150}
\bibitem{footnote} In order to reduce the calculation of $\ln Z_0$ to the calculation of $\langle s_n(\omega)s^*_m(\omega)\rangle$, we first calculate $d(\ln Z_0)/dK$.
\bibitem{footnote1} Indeed, ${\rm Re}\psi_2$ is an eigenfunction with the same energy as $\psi_2$.
\end{thebibliography}
\end{document}